%% file: main.tex
\documentclass[conference]{IEEEtran}
\IEEEoverridecommandlockouts

\usepackage{cite}
\usepackage{amsmath,amssymb,amsfonts}
\usepackage{algorithmic}
\usepackage{graphicx}
\usepackage{textcomp}
\usepackage{xcolor}
\usepackage{caption}
\usepackage{array}
\usepackage{tabu}
\usepackage{booktabs}
\usepackage{multirow}
\usepackage{ragged2e}
\usepackage{booktabs}
\usepackage{threeparttable}
\usepackage{subcaption}
\usepackage{tabu}
\usepackage{lipsum}
\usepackage{mathtools}
\usepackage{cuted}
\usepackage{url}

\def\BibTeX{{\rm B\kern-.05em{\sc i\kern-.025em b}\kern-.08em
    T\kern-.1667em\lower.7ex\hbox{E}\kern-.125emX}}

\usepackage{geometry}
\begin{document}

\newcommand{\fakepar}[1]{{\vspace{4pt}\noindent\textbf{#1}}}

\newcommand{\mmm}[1]{{\color{magenta}{[Maurizio: #1]}}}
\newcommand{\pg}[1]{{\color{purple}{[Paolo: #1]}}}
\newcommand{\mm}[1]{{\color{blue}{[Michela: #1]}}}
\newcommand{\ts}[1]{{\color{brown}{[Tailai: #1]}}}
\newcommand{\DASH}{{\emph{DASH}}}

\title{Modelling Concurrent RTP Flows for End-to-end Predictions of QoS in Real Time Communications\thanks{This work was sponsored by Cisco Systems Inc., and the European Union under the Italian National Recovery and Resilience Plan (NRRP) of NextGenerationEU, partnership on “Telecommunications of the Future”  (PE00000001 - program “RESTART”, Focused Project R4R). This work was supported by the SmartData@PoliTO center on Big Data and Data Science, and the computational resources provided by hpc@polito (http://www.hpc.polito.it).}
}

\author{Tailai Song\IEEEauthorrefmark{1},
Paolo Garza\IEEEauthorrefmark{2},
Michela Meo\IEEEauthorrefmark{1},
Maurizio Matteo Munafò\IEEEauthorrefmark{1}\\
\small{\IEEEauthorrefmark{1}Department of Electronics and Telecommunications Engineering (DET)}\\
\small{\IEEEauthorrefmark{2}Department of Control and Computer Engineering (DAUIN)}\\
\small{Politecnico di Torino, Turin, Italy} \\
\texttt{first.last@polito.it}
}
\maketitle

\input{0_abstract}
\input{1_introduction}
\input{2_problem_context}
\input{3_methodology}
\input{4_experimental_result}

\input{5_conclusion}

\bibliographystyle{ieeetr}
\bibliography{rtc.bib}

\end{document}

%% file: 0_abstract.tex
\begin{abstract}
The Real-time Transport Protocol (RTP)-based real-time communications (RTC) applications, exemplified by video conferencing, have experienced an unparalleled surge in popularity and development in recent years. In pursuit of optimizing their performance, the prediction of Quality of Service (QoS) metrics emerges as a pivotal endeavor, bolstering network monitoring and proactive solutions. However, contemporary approaches are confined to individual RTP flows and metrics, falling short in relationship capture and computational efficiency. To this end, we propose Packet-to-Prediction (P2P), a novel deep learning (DL) framework that hinges on raw packets to simultaneously process concurrent RTP flows and perform end-to-end prediction of multiple QoS metrics. Specifically, we implement a streamlined architecture, namely length-free Transformer with cross and neighbourhood attention, capable of handling an unlimited number of RTP flows, and employ a multi-task learning paradigm to forecast four key metrics in a single shot. Our work is based on extensive traffic collected during real video calls, and conclusively, P2P excels comparative models in both prediction performance and temporal efficiency.
\end{abstract}

\begin{IEEEkeywords}
Real-time communications, RTP, QoS prediction, machine learning, deep learning.
\end{IEEEkeywords}

%% file: 1_introduction.tex
\section{Introduction}
In the wake of unprecedented advancements in recent years, real-time communications (RTC) have become indispensable, offering services such as video-teleconferencing, live streaming, online gaming, and more across professional, recreational, and educational domains. The majority of RTC applications rely on the Real-time Transport Protocol (RTP)~\cite{rfc1889}, while web browsers and mobile devices implement the globally acclaimed open-source framework built atop RTP, known as WebRTC~\cite{loreto2014real}.

As the proliferation of RTC applications persists and is further catalyzed by escalating user demands for optimal Quality of Experience (QoE), there arises a pressing need to develop sophisticated and intelligent technologies aimed at enhancing network performance. In this pursuit, Quality of Service (QoS) monitoring holds paramount importance~\cite{carofiglio2021characterizing, macmillan2021measuring, rao2019analysis}, wherein QoS prediction could assume a pivotal role, facilitating network observability and empowering preemptive solutions~\cite{izima2021survey, morshedi2020survey, kougioumtzidis2022survey}. 

The prediction of QoS metrics has gained considerable attention in research, especially through machine learning (ML) technologies. Focusing on traffic prediction, \cite{song2024bitformer, lee2020perceive} employed deep learning (DL) neural network (NN) to predict bitrate, while \cite{yin2021ant, lv2022lumos} adopted DL and tree-based models for throughput forecasting to enhance adaptive streaming. Authors in \cite{izima2018video, handurukande2011magneto, dinaki2021forecasting} utilized regression models to predict multiple QoS metrics including packet count and jitter. Moreover, packet loss prediction was investigated in \cite{song2023did, giannakou2020machine} using various ML algorithms. Works in \cite{dietmuller2022new, zhang2022delay} targeted network delay prediction based on DL approaches. However, most existing methodologies are tailored solely for a single QoS metric and operate on an individual RTP flow at a time. Intuitively, a plausible solution may entail sequential inference for each individual target (a metric or flow) or parallel processing via multiple threads, yet such approaches prove inefficient, particularly within the context of RTC, where temporal and computational constraints typically exist. On top of that, constrained by a single objective, these methods fail to exploit the underlying connections among flows and metrics to foster the prediction. Additionally, numerous cases necessitate intricate feature engineering and data processing, further exacerbating the problem. 

To address the challenge, we propose Packet-to-Prediction (P2P), an innovative DL framework that operates in an end-to-end manner to predict multiple per-flow QoS metrics directly from raw packets. Capitalizing on the affinity between natural language and network packets, we resort to the Transformer architecture~\cite{vaswani2017attention} to extract features from packet-level information. To tackle the constraint of limited input size, we revamp the architecture with a customized structure named length-free Transformer to manage an unrestricted amount of concurrent RTP flows. We introduce and integrate two types of novel attention mechanisms, cross and neighbourhood attention, to effectively discern both inter- and intra-correlations among flows. Furthermore, we incorporate a multi-task learning block to map the encoded packet features to various QoS metrics. In short, with a single shot, P2P delivers predictions irrespective of the number of flows and metrics. Specifically, our work is grounded in ample RTC traffic gathered during plenty of real video-call sessions, and we formulate a time series problem, utilizing only 128 packets in the past and forecasting 4 key QoS metrics for bitrate, jitter, packet loss, and frames per second (FPS), in future 500-ms time windows. We compare our solution against several ML/DL models proposed in the literature. As a result, P2P not only achieves superior prediction performance but also exhibits significantly reduced time consumption across various computational environments. Additionally, our approach could also serve as a viable venue for comprehending RTC (or even other protocol) traffic "fate", supporting more downstream tasks such as traffic classification and anomaly detection, in addition to or in conjunction with QoS prediction. 

The remainder of this paper is organized as follows: Section~\ref{sec:prob} introduces necessary contexts, including motivations, problem formulation, and the dataset used. Section~\ref{sec:method} outlines our proposed framework and the model evaluation process. Section~\ref{sec:result} presents experimental results, focusing on both prediction performance and time consumption, while offering insights into the model architecture. At last, Section~\ref{sec:con} concludes our work and discusses potential directions for future work.

%% file: 2_problem_context.tex
\section{Problem context}\label{sec:prob}
In this section, we first motivate our work. Subsequently, we formulate the problem and introduce the traffic and dataset employed in our study.

\subsection{Underlying motives}

\subsubsection*{\textbf{Why QoS prediction}}
Understanding network performance is always of utmost significance for stakeholders like end-users and network operators. The prediction of QoS metrics could facilitate multiple network functions: enabling anticipated QoE mapping~\cite{yan2017enabling}, providing critical feedback in advance for congestion control~\cite{carlucci2016analysis}, enhancing dynamic adaptive streaming or encoding~\cite{huang2021dave, liu2023efficient}, and contributing to software-defined networking (SDN) as key indicators for network management like bandwidth allocation~\cite{torres2020sdn}, among others. Although the targeted 500 ms appears to be brief at the first glance, we still envisage its sufficiency and usefulness. Traditional optimization techniques rely on historical observations, which are already outdated, rendering even a slight degree of forward-looking anticipation of QoS metrics advantageous, particularly given the real-time nature of RTC, let alone the possibility of extending our solution to a longer duration.

\subsubsection*{\textbf{Why multiple per-flow QoS metrics}}
On one hand, an RTP flow is determined by a tuple composed of (\textit{IP$_{src}$}, \textit{IP$_{dst}$}, \textit{Port$_{src}$}, \textit{Port$_{dst}$}, \textit{SSRC}, \textit{Type$_{payload}$}). During a video-call session, multiple flows may coexist concurrently, each featuring a distinct media type, sourcing from an individual sender, and traversing diverse network paths to reach the destination. Besides the fact that certain metrics are only derivable separately for each flow, per-flow QoS evaluation could offer nuanced insights into the network performance and conditions from various fine-grained perspectives, allowing networking operations to intervene comprehensively or locally as needed. On the other hand, multiple metrics serve as indicators that could operate either independently or collaboratively, contributing to various network functionalities. Notably, we only target 4 metrics simply because of the unavailability of others in our dataset, and therefore, it is conceivable to extend our approach to predict additional QoS metrics by plugging in supplementary tasks. More importantly, we envisage that our adoption of the multi-task learning scheme could unveil the underlying interdependencies among different metrics, e.g., the occurrence of packet loss may signify a drop in bitrate, to generate enhanced performance. The synchronous optimization of all loss functions could perceive and incorporate the interactions among different tasks, facilitating the prediction to outperform single-task learning, where the impact exerted by other factors remains absent. 

\subsubsection*{\textbf{Why in a single shot}}
Following the previous motivation, unfortunately, present solutions can only address an individual flow or metric each time. To reach the goal of predicting multiple QoS metrics for multiple flows, it leaves us no choice, but to perform sequential or parallel computing, which is feasible yet not efficient. Modern RTC applications often confront temporal and computational constraints, given their real-time nature and restricted control over devices. For instance, certain existing lightweight models prove to be adequate for a single target, but their overall time consumption becomes untenable when applied to multi-targets. Therefore, we aim to devise a model capable of processing and predicting a relatively large amount of flows and metrics without consuming excessive time and extra computational threads, i.e., efficiently accomplishing the task in a single shot. More crucially, notwithstanding the disparities among flows and the predominant interactions entwining packets within a flow, different RTP flows may still share a portion of common origins, network conditions, centralized servers, and bandwidth bottlenecks, manifesting interconnected relationships, which could be taken into account during the modelling of concurrent flows to facilitate predictions, but necessitates the single-shot processing.

\subsubsection*{\textbf{Why packet-level information and Transformer}}
Packets represent the most elemental and granular network units, encapsulating the nuanced characteristics and swiftly changing dynamism of network traffic~\cite{dainotti2008internet}, which enables models to effectively capture traffic patterns for an enhanced performance~\cite{montieri2021packet}. Moreover, it requires minimal effort to extract packet-level data, obviating extra overhead associated with feature extraction and construction. Additionally, we exclusively refer to attributes readily accessible in the RTP header to avoid potential complexities arising from packet encryption. Meanwhile, the sequential nature as well as variable properties inherent in packets bear resemblance to language words (and their tokens), and thus, the Transformer architecture, that holds sway over the Natural Language Processing (NLP) domain and today's Large Language Model (LLM) hype, emerges as a promising candidate for our purpose. Particularly, we employ the Transformer encoder that is frequently utilized for sequence representation~\cite{devlin2018bert, liu2021tera}, harnessing the multi-head attention mechanism with our tailored design to intuitively grasp the intrinsic correlations interweaving packet-level features, RTP flows, and QoS metrics. It is worth noting that while certain variants of Transformer can indeed handle unlimited input~\cite{bertsch2024unlimiformer}, they are not applicable to our specific scenario because of the fundamental difference in our primary objectives. Additionally, Transformers have also been applied for QoS prediction in other domains~\cite{hameed2022toward, kumar2024tpmcf}, but they still predict only a single metric at a time.

\subsection{Problem formulation}
At a time instant $t$, our objective is to predict $M$ QoS metrics for $L$ flows (a total of $M \times L$ metrics) in the subsequent time window of duration $\Delta t$, based on $n$ packets for each flow out of all the preceding $N$ packets. 

In our work, we aim at a time span of $\Delta t=500$ ms, during which $M=4$ metrics are considered: \textit{1)} Bitrate ($R_{bps}$) --- the total size of all packets for a certain flow transmitted over the given duration; \textit{2)} Average jitter ($\overline{R}_{jitter}$) --- the mean value of all inter-arrival jitters, each calculated in accordance with the RTP specifications~\cite{schulzrinne2003rfc3550}\footnote{It differs from the simple definition of network jitter (the variance of inter-arrival time) and requires the clock rate (packet generation frequency) of a particular payload type, which is accessible in our video conferencing logs.}; \textit{3)} FPS ($R_{fps}$) --- the number of frames per second for video flows, derived from the count of unique RTP timestamps; \textit{4)} Loss condition ($y_{loss}$) --- a binary class label indicating the presence of packet loss (0 -- lossless, 1 -- lossy), determined by the absence of sequence numbers. To sum up, we formulate a combination of 3 regression problems and 1 classification problem.

Furthermore, we refer to a total of $N=2048$ packets in the past and define an active flow accordingly: an RTP flow is deemed active if (1) it comprises at least 128 packets in the preceding 2048-packet range and (2) its last (latest) packet is within 1 second away from time $t$ (the start of the predicted window). This criterion is akin to the reality where predictions are unnecessary for inactive or interrupted flows. Hence, $L$ denotes the total number of active flows, each accompanied with a fixed quantity of $n=128$ packets for predicting corresponding metrics. Notably, the aforementioned parameters of $\Delta t$ and $n$ are modifiable, and our preliminary results attest to consistent performance, which we elaborate on in future works. The initial choice of a 500-ms window is driven by the need for frequent updates of QoS metrics, while the selection of 128 packets is motivated by the desire to minimize packet usage, thereby enhancing efficiency without significant loss of information. On top of that, we employ 6 elements of each RTP packet as features: absolute inter-arrival time (relative to the preceding packet), relative inter-arrival time (compared to the first packet in the 128-packet series), packet size, RTP timestamp, RTP marker, and sequence number difference (between adjacent packets). In essence, we expect to encompass the influence exerted by multifarious facets from temporal, spatial, and RTP-related components.

\subsection{Traffic \& Dataset}
The traffic employed in our work is collected during 71 independent real video conferences (roughly 69 hours). The comprehensiveness and diversity of our data are reflected in four key aspects: 2–6 participants located across Europe, connectivity via WiFi, mobile, or Ethernet, a collection period spanning several months from April 2020 to January 2021, and the streaming of various media content, including low/high-quality video, audio, and screen sharing. We utilize two RTC applications, \textit{Jitsi Meet}\footnote{An open source platform, \url{https://meet.jit.si/}.} and \textit{Webex}\footnote{A commercial application, \url{https://www.webex.com/}.}, capturing traffic on client sides and storing in \textit{pcap} format. Particularly, we only focus on incoming streams to predict QoS metrics of traffic originating from all sources as well as influenced by entire network links and nodes.

According to the problem formulation, we construct our dataset by creating a time series of consecutive time windows for each session, i.e., a \textit{pcap} file. For each window, we identify active flows in the past, and then for each flow, we calculate corresponding metrics in the window. Consequently, a time window yields one or more samples ($L$ active flows), each composed of the 4 QoS metrics as prediction targets and the previous 128 packets containing the aforementioned 6 elements as features. In total, 1,631,436 samples are generated in our dataset. Importantly, on average, each time window contains 3.4 coexisting active flows, with a maximum of 11, and only 10\% of windows feature a single flow, which further underscores our motivation to ensure efficient processing. 

\begin{figure*}[t]
\centering
\includegraphics[scale=0.3]{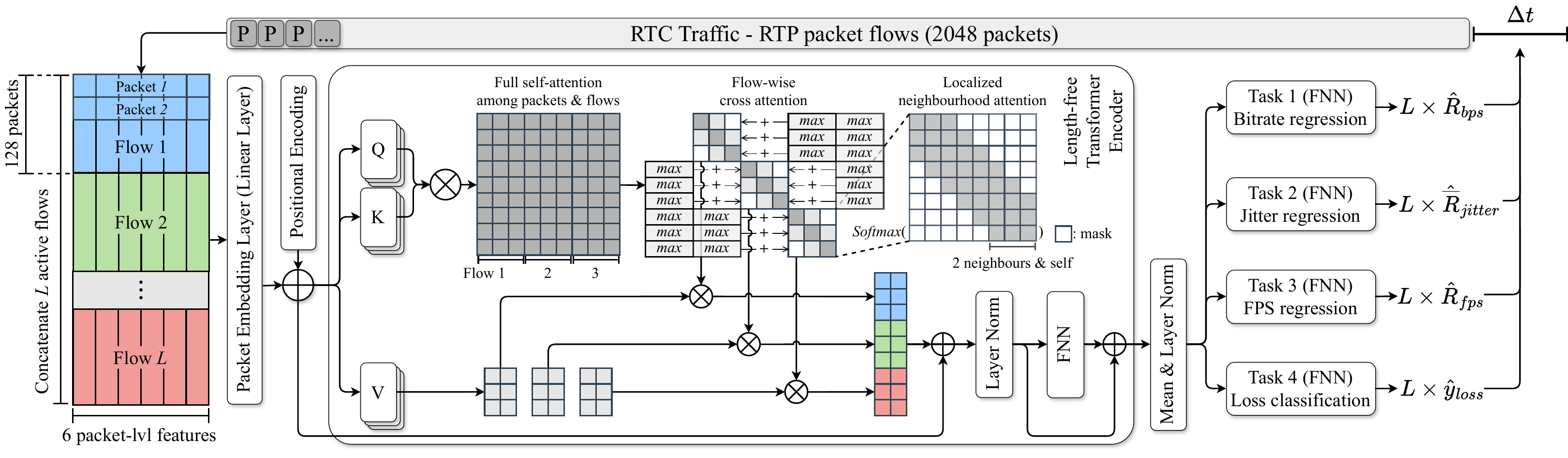}
\caption{P2P: workflow \& model architecture.}
\label{fig:arch}
\end{figure*}

%% file: 3_methodology.tex
\section{Methodology}\label{sec:method}
In this section, we delineate the proposed model, and then illustrate the performance evaluation process.

\subsection{Proposed model}
As illustrated in Figure~\ref{fig:arch}, P2P receives packets from all active flows, extracting features with a customized Transformer encoder, and predicts multiple metrics through parallel tasks. 

\subsubsection*{\textbf{Packet processing}}
Initially, we extract from each session active flows along with their respective packets, organized in chronological order. Then, all flows, i.e., every 128 packets with 6 packet-level features, are concatenated to form a $128 \cdot L \times 6$ matrix as input. Subsequently, P2P incorporates a packet embedding layer (a linear layer) to convert the input into embedded features (a $128 \cdot L \times N_{embedding}$ matrix), mapping packet attributes to their latent embeddings, and thereby enriching the traffic features. Adhering to the good practice of the vanilla Transformer model, the positional encoding is then superimposed on top of embedded features.

\subsubsection*{\textbf{Length-free Transformer}}
The conventional Transformer is bounded to a fixed input length, and thus cannot handle an indeterminate number of flows (variable $L$). To overcome this limitation, we meticulously devise a dual attention mechanism to make the Transformer length-free. 

Besides the fundamental role of modelling packet sequences within and among flows, the first part, named flow-wise cross attention, is primarily responsible for dealing with variable flows. The core of the Transformer model (i.e., the multi-head self-attention) constitutes a series of computations for the query ($Q$), key ($K$), and value ($V$) generated from the input features. The initial step forms the attention matrix by multiplying the query and key ($W=QK^T$). As exemplified by the full (regular) self-attention in Figure~\ref{fig:arch} (the leftmost greyish matrix), it results in a $9 \times 9$ square matrix ($W_{9 \times 9}$) if the input includes 3 flows, each having 3 packets (i.e., $nL \times nL; L=3, n=N_{embedding}=3$). Let $W_{9 \times 9}(i,j)$ denote the element in row $i$ and column $j$ of the matrix, then $W_{9 \times 9}(1,1)$ and $W_{9 \times 9}(1,2)$ represent the attentions paid by the $1^{st}$ packet in flow 1 to itself and to the $2^{nd}$ packet also in flow 1, respectively, and thereby $W_{9 \times 9}(1$:$3,1$:$3)$ are the attentions acknowledged within flow 1. In contrast, $W_{9 \times 9}(1,4$:$6)$ and $W_{9 \times 9}(1,7$:$9)$ are the attentions paid by the $1^{st}$ packet in flow 1 to packets belong to flows 2 and 3, respectively. Similarly, the elements in $W_{9 \times 9}(1$:$3,4$:$9)$ are attentions paid by flow 1 to other flows. In this vein, as elucidated by Figure~\ref{fig:arch} (the collection of matrices on the right of the full self-attention), the sub-matrices on the diagonal represent intra-correlations within individual flows, while other square matrices stand for inter-correlations among flows. In other words, by considering all flows as a whole during input but treating them separately when computing attentions, we can scale up to a theoretically infinite amount of concurrent RTP flows.

On top of that, in addition to the primary intra-correlations among packets within the same flow, we also envision potential relationships among concomitant flows. To this end, we introduce flow-wise cross attention, which focuses on the diagonal attention matrices augmented by refined information from other flows, as depicted in Figure~\ref{fig:arch}. Each row in an off-diagonal sub-matrices represents the attentions paid by a certain packet in the current flow to all packets in another flow. In order to bring in the inter-correlations from other flows, we first apply max-pooling to such rows, capturing the most significant impact, i.e., calculating the maximum value of attentions paid by a packet to a certain flow of its extrinsic ones (a row in an off-diagonal matrix), and summing them up for all other flows. Subsequently, we add the resultant summation to each of the attention values within the current flow paid by the corresponding packet. For example, $W_{(9 \times 9)}(1, 1$:$3) \leftarrow W_{(9 \times 9)}(1, 1$:$3) + (\text{max}(W_{(9 \times 9)}(1, 4$:$6))+\text{max}(W_{(9 \times 9)}(1, 7$:$9)))$. In essence, the flow-wise cross-attention confines and prioritizes the packet interactions to those localized within the same flow while still accounting for a certain degree of influence from other flows.

Moreover, we further modify each sub-attention matrix on the diagonal, implementing localized neighbourhood attention, inspired by \cite{hassani2023neighborhood}. In our work, $n=128$ packets per flow are considered. We argue that the correlations are weak between packets that are temporally distant from each other, while substantial between adjacent ones. Because of the rapidly oscillating network dynamics, patterns manifested by, for example, the first and the last packets with seconds in between, might differ significantly, demonstrating minimal relationship. Therefore, instead of relying on the NN to learn such a characteristic, we intentionally localize the attention paid by a packet only to itself as well as its $k$ nearest neighbours within a flow. As portrayed by the last zoomed-in matrix in Figure~\ref{fig:arch} (the rightmost $8 \times 8$ example matrix with grey and white mixed), such a goal can be achieved by applying masks to needless elements (distant attentions). In the case of an 8-packet flow (sub-matrix) with a neighbourhood degree of $k=2$, we only retain the diagonal array, where at least 3 attention scores are reserved on a row (attention paid by an edge packet to itself and 2 packets on left or right) and at most 5 are reserved (itself and the 2 packets on either side). This method assures that the model focuses on closely related packets, eliminating noises introduced by irrelevant packets and enhancing the accuracy of the attention mechanism.

Consequently, we finalize the attention scores by combining the aforementioned 2 operations, and only retaining the diagonal sub-matrices for individual flows, each multiplying its corresponding value matrix ($V$) to derive the final output. Afterwards, we concatenate all outputs, proceeding with the remaining layers of the Transformer encoder to produce a $128L \times N_{embedding}$ matrix, i.e., encoded features. By averaging along the second dimension, we distil the encoded features, and by performing layer normalization, we stabilize the values and conclude the feature extraction phase.

\subsubsection*{\textbf{Multi-task learning}}
The output of the previous stage is a $128L \times 1$ vector, wherein every 128 entries represent the ultimate encoded features for the 128 packets of a flow (one out of $L$ flows). To predict the 4 target QoS metrics, we employ a multi-task learning paradigm with 4 parallel regression/classification blocks, each comprising a feedforward neural network (FNN). Each set of the encoded features (a $128 \times 1$ vector) of an individual flow is fed into a block to forecast the corresponding metric, resulting in $4L$ predictions for all active flows. To facilitate the training and optimize the performance, the regression problems are associated with Mean Absolute Error (MAE) loss, whereas the classification one involves weighted Binary Cross Entropy (BCE) loss with a higher weight assigned to the minority class (1, lossy time windows) to tackle the class imbalance issue, given that only 1.7\% of samples are lossy. More importantly, we implement learnable weights~\cite{kendall2018multi} to automatically ascertain the importance of each task, systematically melding losses produced by different blocks. At last, we enumerate the implementation details regarding hyperparameters in Table~\ref{tab:para}

\begin{table}[t]
\centering
\scriptsize
\captionsetup{font=normalsize}
\caption{Implementation detail of \textit{DeX}}
\label{tab:para}
\begin{threeparttable}
\begin{tabular}{l|l} 
\toprule
\textbf{Parameter} & \textbf{Value} \\ 
\midrule
Learning rate{*}, $\eta$ & $10^{-3}$ \\
Size of feature embedding, $N_{\text{embedding}}$ & 32 \\
Neighbourhood degree, $k$ & 32 \\
Number of heads & 8 \\
Number of neurons for the FNN in encoder & 512 \\
Activation function in encoder & \textit{LeakyReLU}~\cite{xu2020reluplex} \\
Number of layers for multi-task learning blocks & 2 \\
Number of neurons of the $1^{
\text{st}}$ layer for a task & 64 \\
Number of neurons of the $2^{\text{nd}}$ layer for a task & 1 \\
Activation in multi-task learning blocks & \textit{LeakyReLU} \\
Training optimizer & \textit{Adam}~\cite{kingma2014adam} \\
Batch size & 8 \\
\bottomrule
\end{tabular}
\begin{tablenotes}\scriptsize
\item[*] A decay of 1 order of magnitude for every 2 epochs is adopted.
\end{tablenotes}
\end{threeparttable}
\end{table}

\subsection{Model evaluation}
The 3 regression problems for bitrate, jitter, and FPS prediction with packet-level features can be also framed as univariate time series forecasting problems using historical samples as features. As for the classification problem for loss prediction, we refer to \cite{song2023did}, formulating a multivariate time series problem and creating traffic statistics on an aggregated manner for all packets in a time window. In light of this, we also refer to 4 common ML/DL models to compare P2P: Random Forest (RF)~\cite{breiman2001random}, eXtreme Gradient Boosting (XGB)~\cite{chen2015xgboost}, Multi Layer Perceptron (MLP)~\cite{goodfellow2016deep}, and Long Short-Term Memory (LSTM)~\cite{hochreiter1997long}. In our dataset, the $N=2048$ packets under consideration in the past translate to an average time span of approximately 7.8 s, and we thereby opt for an equivalent duration of 8 s (16-time windows) for comparative models to make predictions, e.g., the bitrate value in each of the 16 preceding windows for future bitrate prediction. Additionally, we also implement another packet-level model for comparison, which leverages LSTM for feature extraction, utilizing two layers of LSTM cells and channelling the last hidden state of the second layer into subsequent multi-task learning blocks. Notably, this model is still unable to handle concurrent flows simultaneously, but it processes them sequentially. It is important to emphasize that we intentionally limit our consideration to the aforementioned comparative models for two main reasons: 1) most existing works indeed employ common algorithms, and it is nearly impossible to identify a unified approach applicable to all the considered QoS metrics simultaneously, and 2) many state-of-the-art time-series models are not only incompatible but also feature complex architectures that are time-intensive, which inherently contradicts our primary objective.

Furthermore, instead of shuffling the entire corpus of available data, we randomly split the 71 video-call sessions into 3 independent subsets of 50, 10, and 11, to create training (1,218,164 samples), validation (159,132), and test (254,140) datasets. As a result, our model is trained on traffic exhibiting unique conditions that differ from those in other datasets in terms of collection time, location, connectivity, etc., intending to prevent data infiltration and to develop a generalized solution. Note that each sample corresponds to an individual flow within a time window, and while all active flows in a given window are processed simultaneously in a single pass by P2P, they are handled separately by other comparative models. Consequently, the number of available time windows is 345,812 for training, 51,446 for validation, and 78,750 for the test dataset. This means that each training sample (time window) for P2P may encompass a varying number of RTP flows. To evaluate the performance, we gauge 4 metrics for regression problems: Root Mean Squared Error (RMSE), Mean Absolute Error (MAE), Mean Absolute Percentage Error (MAPE), and the coefficient of determination (R$^2$ score). Concerning the classification problem for loss condition, not all metrics are representative owing to the class imbalance, e.g., precision is biased towards the majority class. Therefore, we refer to class recalls and the F1-score to assess the performance of individual classes and the entirety.

%% file: 4_experimental_result.tex
\section{Experimental result}\label{sec:result}
In this section, we present the experimental outcomes for all QoS metrics and models. Then, we elaborate on the temporal efficiency of the models.

\subsection{Prediction performance}

\begin{table}
\centering
\caption{Experimental results of all models.}
\label{tab:result}
\scriptsize
\begin{threeparttable}
\begin{tabular}{c|c||>{\centering}p{0.5cm} >{\centering}p{0.5cm} >{\centering}p{0.5cm} >{\centering}p{0.5cm} >{\centering}p{0.5cm} c}
\toprule
\multicolumn{2}{c||}{Problem} & \multicolumn{4}{c|}{Time series} & \multicolumn{2}{c}{Packet-level} \\ 
\midrule
\multicolumn{2}{c||}{Model} & RF & XGB & MLP & \multicolumn{1}{c|}{LSTM} & LSTM & P2P \\ 
\midrule
\multirow{4}{0.09\linewidth}{\hspace{0pt}\Centering{}\begin{tabular}[c]{@{}c@{}}Bitrate\\{[}Mbps]\end{tabular}} & RMSE $\downarrow$ & 0.100 & 0.100 & 0.100 & 0.103 & 0.132 & \textbf{0.098} \\
 & MAE $\downarrow$ & 0.031 & 0.032 & 0.030 & 0.032 & 0.051 & \textbf{0.029} \\
 & MAPE $\downarrow$ & 16.65\% & 17.83\% & 12.40\% & 14.63\% & 19.30\% & \textbf{9.77\%} \\
 & R$^2$ $\uparrow$ & 0.965 & 0.966 & 0.965 & 0.964 & 0.939 & \textbf{0.967} \\ 
\midrule
\multirow{4}{0.09\linewidth}{\hspace{0pt}\Centering{}\begin{tabular}[c]{@{}c@{}}Jitter\\{[}ms]\end{tabular}} & RMSE $\downarrow$ & \textbf{1.583} & 1.675 & 1.697 & 1.586 & 1.875 & 1.648 \\
 & MAE $\downarrow$ & 0.762 & 0.775 & 0.853 & \textbf{0.753} & 0.895 & 0.795 \\
 & MAPE $\downarrow$ & 14.56\% & 15.16\% & 17.24\% & \textbf{14.05\%} & 16.82\% & 14.97\% \\
 & R$^2$ $\uparrow$ & \textbf{0.860} & 0.844 & 0.840 & \textbf{0.860} & 0.804 & 0.849 \\ 
\midrule
\multirow{4}{0.09\linewidth}{\hspace{0pt}\Centering{}\begin{tabular}[c]{@{}c@{}}FPS\\{[}fps]\end{tabular}} & RMSE $\downarrow$ & 2.087 & 2.733 & 2.454 & 2.606 & 1.840 & \textbf{1.808} \\
 & MAE $\downarrow$ & 0.927 & 1.053 & 1.026 & 1.090 & 0.842 & \textbf{0.800} \\
 & MAPE $\downarrow$ & 6.39\% & 7.60\% & 7.15\% & 7.65\% & 6.09\% & \textbf{5.67\%} \\
 & R$^2$ $\uparrow$ & 0.959 & 0.929 & 0.943 & 0.936 & 0.966 & \textbf{0.967} \\ 
\midrule
\multirow{3}{0.09\linewidth}{\hspace{0pt}\Centering{}\begin{tabular}[c]{@{}c@{}}Loss\\{[}-]\end{tabular}} & Recall-0 $\uparrow$ & 0.856 & \textbf{0.968} & 0.794 & 0.777 & 0.953 & 0.950 \\
 & Recall-1 $\uparrow$ & \textbf{0.622} & 0.415 & 0.183 & 0.267 & 0.517 & 0.574 \\
 & F1-score $\uparrow$ & 0.540 & 0.638 & 0.463 & 0.457 & 0.631 & \textbf{0.641} \\
\bottomrule
\end{tabular}
\begin{tablenotes}\scriptsize
\item[*] $\uparrow$: the higher the better, $\downarrow$: the lower the better.
\end{tablenotes}
\end{threeparttable}
\end{table}

\begin{figure*}[t]
\centering
\includegraphics[scale=1.05]{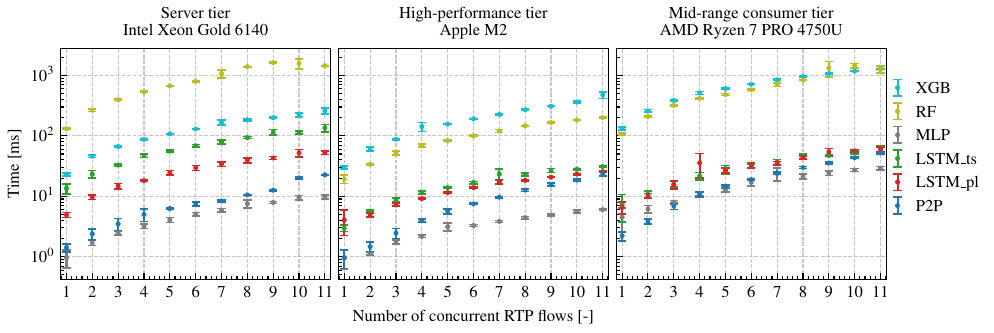}
\caption{Time consumed by each model for various number of flows in different CPU environments.}
\label{fig:ts}
\end{figure*}

Table~\ref{tab:result} showcases the performance evaluation of all models. In general, P2P outperforms comparative models in most cases. In terms of bitrate prediction, P2P yields superior performance, standing out as the sole model with a MAPE lower than 10\%, while the disparities in other metrics are minor. As for predicting average jitter, time-series models generally outstrip packet-level ones, likely due to the explicit inclusion of historical jitter values in the time-series features, enabling models to catch recent jitter variations and thereby operate autoregressively. In contrast, we deliberately exclude preceding jitter values from packet-level features to avoid extra overhead as jitter is computed iteratively for each packet, and to depend on the models to ascertain the intrinsic correlations between jitters and packet features. While we insist on such a strategy since the primary objective is to boost efficiency, P2P still delivers comparable results with a slight performance decline. Regarding FPS prediction, the superiority of P2P is even more pronounced, evidenced by being the only model with a MAPE below 6\%. Interestingly, both packet-level models demonstrate prowess compared to time series models, boasting decent RMSE and MAE with magnitude-level improvement, and potentially illustrating the inherent capability of packet-level information in capturing FPS evolution. Moving forward to loss conditions, P2P does not attain the best class recalls. Nonetheless, the decline in recall concerning class 0 is trivial with respect to the top rank (XGB), whose class-1 recall is, however, roughly 16\% lower, and while RF with the highest class-1 recall transcends P2P by around 5\%, it incurs excessive misclassifications for lossless windows, leading to subpar recall for class 0. In contrast, P2P strikes a balance between both classes, adeptly identifying the majority of lossy time windows without penalizing lossless ones, and thereby highlighting the premier F1-score. Additionally, evaluating the LSTM-based packet-level model can also be viewed as an ablation test, highlighting the prowess of the Transformer encoder for feature extraction.

\subsection{Time consumption}
Importantly, although P2P exhibits enhanced outcomes, the performance comparison analysis does not hold paramount significance, as our primary focus lies on efficiency through the management of concurrent RTP flows and multiple QoS metrics. To this end, we also evaluate the temporal proficiency of P2P, by comparing against other considered models. We examine the time consumption across 3 different computational environments, each featuring varying levels of CPU devoid of GPU acceleration, to evaluate the time required to predict all 4 QoS metrics for a range of concurrent RTP flows from 1 to 11.

Figure~\ref{fig:ts} presents the experimental findings. As shown, all models experience a relatively linear growth (note the logarithm scale) as the flow quantity increases, with P2P exhibiting superior efficiency by consuming the second least time across different computational environments. The results are generally stratified, with different patterns emerging: the server-tier CPU manifests distinctly varied time consumption among models, while for the other two CPUs, XGB and RF incur excessive time costs, reaching up to 1 second, and other models yield similar performance with more reasonable time consumed. In particular, both server-tier and high-performance CPUs exhibit similar behavior, where P2P processes a single flow in mere milliseconds, and up to roughly 25 ms for 11 flows. As for the least powerful CPU, P2P needs approximately 50 ms in the worst-case for 11 flows, still leaving more than a quarter of the available time in the predicted 500-ms window, and even outpacing the time required by XGB or RF to address a single flow. Notably, MLP demonstrates the fastest processing time regardless of CPUs thanks to its inherently streamlined architecture, and both LSTM models produce time consumption similar to P2P on high-performance and mid-range tier CPUs, especially with a large number of concurrent flows. However, their overall quantitative performance remains subpar. Although P2P does not attain the highest efficiency, and its improvements over certain models are modest due to the quadratic complexity of the attention mechanism, it strikes a balance by delivering optimal prediction performance without excessive time consumption, not to mention the potentiality of implementing linear-complexity Transformer~\cite{katharopoulos2020transformers}.

\begin{figure}[t]
\centering
\includegraphics[scale=1.2]{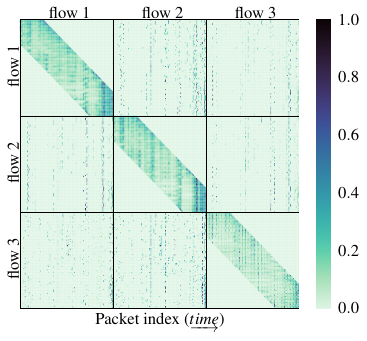}
\caption{Normalized attention matrix with max pooling highlighted for a random 3-flow sample (time window).}
\label{fig:attn}
\end{figure}

\begin{table*}
\centering
\caption{Experimental results for the original Transformer model.}
\scriptsize
\label{tab:ab}
\begin{tabular}{llll|llll|llll|lll} 
\toprule
\multicolumn{4}{c|}{Bitrate [Mbps]} & \multicolumn{4}{c|}{Jitter [ms]} & \multicolumn{4}{c|}{Frame per second [fps]} & \multicolumn{3}{c}{Loss [-]} \\
RMSE & MAE & MAPE & R$^2$ & RMSE & MAE & MAPE & R$^2$ & RMSE & MAE & MAPE & R$^2$ & Recall-0 & Recall-1 & F1-score \\ 
\midrule
0.099 & 0.029 & 9.86\% & 0.966 & 1.703 & 0.813 & 15.60\% & 0.838 & 1.883 & 0.821 & 5.93\% & 0.960 & 0.948 & 0.585 & 0.640 \\
\bottomrule
\end{tabular}
\end{table*}

Overall, the reduced time consumption of P2P stems from its capacity to share features across various tasks and simultaneously handle everything in one single pass, thereby substantially reducing computational overhead, whereas most other models deal with one flow/metric at a time to make predictions sequentially (e.g., 44 iterations of computation for 4 metrics and 11 flows). In essence, the time conserved by P2P empowers the execution of predictions to operate in a real-time manner, offering a greater degree of flexibility to implement network optimization measures promptly and timely. Noteworthily, the performance of P2P herein could be further enhanced, as we do not incorporate any potential optimizations, such as well-known NN pruning techniques~\cite{liang2021pruning} and other effective Transformer structures~\cite{10.1145/3530811}. Meanwhile, time-series models might be even less efficient due to the feature construction process not being factored in.

\subsection{Insights into attention mechanism}
In order to substantiate the contribution of our customized attention mechanism, we showcase an example of attention matrix for a randomly selected sample comprising 3 concurrent flows, as portrayed by the heatmap in Figure~\ref{fig:attn}. The diagonal $128 \times 128$ sub-matrices correspond to the final attention scores for each flow, while the off-diagonal matrices reflect the corresponding maximum attention values allocated to non-self flows\footnote{The presence of multiple maximum values per row for a given flow arises from the independent computation of max values for each attention head, and in our case, there are 8 heads.}.

Distinctively, our design for localized neighborhood attention results in belt-like attention scores along the diagonal. Furthermore, temporally adjacent packets demonstrate stronger attentions, as indicated by the deeper color, which aligns with our expectations --- recent packets are likely to capture the most current traffic dynamics. Although all 3 flows display similar overall patterns, marked differences can still be observed. For instance, the max poolings are more sparsely distributed for flow 3 (the two lowest sub-blocks on the left), whereas they are concentrated on specific packets in the other flows. This demonstrates that P2P can not only grasp intra-flow correlations (as evidenced by non-null values) but also discern distinguishing patterns across individual flows.

To further consolidate the effectiveness of the cross and neighbourhood attention, we conduct an ablation test, substituting our customized feature extraction block with the vanilla Transformer encoder using a standard attention mechanism. Notice that it can only process parallel RTP flows sequentially, while all other architectures as well as configurations remain unaltered. The results, shown in Table~\ref{tab:ab}, indicate that although the original Transformer model still performs reasonably well compared to other ML/DL models, highlighting its innate capability of modelling packet sequences, it falls short of surpassing our customized design, which, in turn, underscores the significance of capturing both inter- and intra-flow correlations.

%% file: 5_conclusion.tex
\section{Conclusion}\label{sec:con}
In this paper, we propose Packet-to-Prediction, a Transformer-based DL framework that can handle an unlimited number of concurrent RTP flows to simultaneously forecast 4 key QoS metrics in a single run. Moreover, P2P leverages the finest granular features from raw packets to perform end-to-end predictions. To surmount the Transformer's restriction on bounded input and to introduce inter-flow correlations, we introduce a customized architecture, termed cross and neighbourhood attention, which localizes attention scores within an individual flow and an entry's neighbourhood. Afterwards, we implement a multi-task learning paradigm for the sake of parallel metric predictions. Our work is rooted in massive traffic collected during real video-teleconferencing sessions under diverse conditions. We benchmark P2P with multiple ML/DL technologies, and consequently, our solution presents preeminent performance, especially regarding time efficiency. As a result, we envision that P2P is able to empower a more effective and timely network management system, as it delivers accurate, find-grained predictions with comparatively sufficient lead time, transcending traditional frameworks that rely solely on historical observations. We acknowledge that it is currently challenging to assess the tangible performance gains of P2P in real-world applications, which, however, is beyond the scope of this work and can be further explored. Future works could also consist of an investigation into more QoS indicators and their applicability to other tasks.